# A case for FDI in Multi-Brand Retail in India

Jatin Prasad

Research Scholar

Rajasthan College, Jaipur

Dr Jyoti Singh

Associate Professor

Rajasthan University, Jaipur

**Abstract**

India is ranked as the third most attractive nation for retail investment among emerging markets and many MNCs have been looking for the potential benefits to be taken from it. The development of organized retail has the potential of generating employment, improvement in technology, development of real estate etc. On the other hand critics of the FDI feel that allowing FDI would jeopardize the unorganized retail sector and would not only adversely affect the small retailers and consumers but will give rise to monopolies of large corporate houses also, which can adversely affect the pricing and availability of goods. A case for the prospects for the same is discussed in this paper.

**Keywords**: FDI, multi-brand retail, Wal-Mart

## 1. Introduction

The retail industry in India is divided into two sectors: organized and unorganized sectors. Since 1991, when economy was liberalized, organized retail has grown exponentially because of burgeoning purchasing power of Indian middle class and the opening up of the foreign direct investment ("FDI") to it. As per the present regulations 100% FDI is allowed in wholesale trading and 51% is allowed in single-brand retailing. The FDI in multi-brand retailing has been just allowed and so the question, if opening up of FDI in multi-brand retail is a boon or a curse, is opened up.

## 2. The Foreign Direct Investment

FDI is defined as "investment made to acquire lasting interest in enterprises operating outside of the economy of the investor. It is a long-term relationship between the investor and the recipient entity." (www.unctad.org)

For FDI there is a need of a 10% holding or greater. Most FDI ends up being 100% ownership by a Multi-National Corporation (MNC). In the past twenty years, FDI has increased dramatically and has become the most common type of capital flow across the borders of world economies.

There are two matching ways and concepts of measuring the FDI:

- It is a particular form of the capital flow across international borders giving rise to a

- particular form of international assets for the home countries. To be specific, the value of holdings in corporations, controlled by a home country resident or in which a home country resident holds a certain share of the voting rights.
- It is a set of economic activities or operations carried out in a host country by firms controlled or partly controlled by firms in some other (home) country. These activities are, for example, employment, production, sales, the use and purchase of fixed capital and intermediate goods, and the carrying out of R&D.

FDI is not only beneficial to individuals of the society but it is spread throughout the economy also via the theory of the multiplier effect, which ultimately boosts the economy of a country raising its standard of living.

For the most part economists welcome the increase of FDI to developing economies. It brings capital needed for economic development in the country in a way that is not as risky as borrowing from overseas. It may also bring a range of additional benefits.

The benefits of FDI arise from:

- The important benefit is from the trickle-down effect of the infusion of capital in the national economy of the country.
- As MNEs typically possess state-of-the-art technology and as potential host countries often lack the resources for R&D to acquire the latest technology, so they may acquire it from MNEs by FDI.
- The MNE can train local people of host countries with its management skills to occupy managerial, financial and technical positions within the MNE, which eventually become generally available within the host country.
- The jobs are created directly and indirectly through the multiplier effect. The indirect employment effects may equal, and sometimes exceed, the direct effects. FDI sometimes also eliminates jobs through displacement of local firms, but the net effect is generally additional employment within the host economy. (Hill, 2005, p. 246)
- FDI has a number of BOP benefits for a host country:
- Due to an investment by MNE, the capital account registers a credit - a one-off credit.
- The current account will register credits, if the FDI is a substitute for imports
- If the MNE uses the investment for exports, again it will be the current account credit.

Potential costs of FDI are:

- It is expected that a well-financed MNE with managerial expertise can drive existing local companies out of

business. It may also inhibit the growth of infant industry.

- The continuing outflow of earnings from the FDI and imports of components as part of total production process will be shown as debits on the capital account.
- An emotional reaction can be evoked by the presence of an MNE in a host country to apparent economic domination and loss of national identity.

3. **A case for FDI in multi-brand retail in India**

There is a lot of talk about allowing FDI in multi-brand retail in India. There are still some prohibitions for FDI in multi brand retail trade in India. The country, however, has recently allowed up to 51 per cent FDI in multi brand retail like single brand.

The Wal-Mart has been thinking to enter the multi-brand retail in India with FDI. "It is for the government to deal with and as and when they come out with the conditions for allowing such investments. The final decision will depend on government's future stand on its investment policy in multi brand retail. We expect the government to consider allowing at least 50 per cent FDI in multi-brand retail and then there are other terms and conditions which too will be considered for shaping the future of Wal-Mart in India," said Wal-Mart country head in India.

Wal-Mart has currently joint venture with Bharti that has six large format wholesale stores. It supplies to 130 Easy Day retail outlets owned by Bharti Retail. Although Wal-Mart is pleased with their journey so far with Bharti, but it has been planning to open 10 stores this year and a similar number next year, mostly in tier-II and tier-III cities of south and west India. The reason behind this is that they believe there is more demand for multi brand from consumers in India who also hold a large potential.

Wal-Mart India aims to establish farm linkages and work with the farmer to improve productivity so as to establish linkages with its 1,000 plus suppliers in the country.

The future of Wal-Mart in India is quite prospective. There is a boom in Indian retail scenario and Indian retailers are performing well across the board. In the organized retail trade current profits are good and the future appears to be even more alluring. The savings are less today by an average urban consumer than it were a few years ago and more importantly, expenditure of income is on a wider array of goods than it was earlier. It is the expansionist mood of the current market as consumers are willing to lap up better and newer brands. By 2012, the retails markets 16% account is to be captured by organized retailers, which would be quite up from 4% of today. The India's retail trade size is estimated at $206 billion and it has been growing at five per cent per annum.

Worldwide, the business of Wal-Mart revolves around a very efficient supply chain, with savings in each aspect of that supply chain. It's not just about negotiation to get better prices with the suppliers; it is actually to remove inefficiencies in the supply chain.

"Our relationship with India is broader. We have a sourcing relationship; we buy a lot of merchandise out of India. Again, I think the wholesale opportunity is big. My experience over the last few years has been in the wholesale business in the US serving small business. I believe there is a role for us to play here serving *kiranas* and hotels and restaurants and other people and we'll work on those things. Also, we are trying to create the retail infrastructure as we serve the Bharti stores with cold chain, with a relationship with farmers and others, so that things are ready if and when it does present an opportunity for us to open stores directly.

We are long-term thinkers. I think if you look back at the history of Wal-Mart's business, not only outside the US, but even in the US, we like to start slowly and learn the business and make our mistakes and fix them and once it is time, the government, customers and society is ready, we can do things faster if that is the case. But if not, we will just continue to go along as we are and have a long-term view" said Doug McMillon, President and Chief Executive Officer of Wal-Mart International.

## 4. Conclusion

With the case of Wal-Mart, it can be said that FDI in multi-brand retail in India should be given a serious thought and a gradual opening up must be made possible. In spite of country wide speculation on the plight of small retailers, India must take a lesson from China, where organized and unorganized retail is co-existing and growing together. With allow of FDI in multi brand retails, local enterprises of India will potentially receive an up-gradation with the import of advanced technological and logistics management expertise from the foreign entities to improve its infrastructure, access sophisticated technologies and generate employment for those keen to work in this sector. The FDI would lead to a more comprehensive integration of India into the worldwide market and, as such, it is imperative for the government to promote this sector for the overall economic development and social welfare of the country. If done in the right manner, it can prove to be a boon and not a curse.